\documentclass[11pt]{article}
\usepackage{graphicx}

\def\be{\begin{equation}}
\def\ee{\end{equation}}
\def\ba{\begin{eqnarray}}
\def\ea{\end{eqnarray}}

\def\12{{1\over 2}}

\def\msun{M_\odot}

\def\etal{{\it et~al.~}}
\def\ltsima{$\; \buildrel < \over \sim \;$}
\def\simlt{\lower.5ex\hbox{\ltsima}}
\def\gtsima{$\; \buildrel > \over \sim \;$}
\def\simgt{\lower.5ex\hbox{\gtsima}}

\begin{document}

\title{\bf Instabilities in the Ionization Zones \\ Around the First Stars\footnote{This paper is published
in Astronomy Reports, 2012, Vol. 56, No. 7, pp. 564-571.}}
\author{E.~O.~Vasiliev$^{1,4}$\thanks{eugstar@mail.ru}, E. I. Vorobyov$^{1,2}$, A.~O.~Razoumov$^3$,
Yu.~A.~Shchekinov$^{4,5}$ \\
\it $^1$Institute of Physics, Southern Federal University, Rostov-on-Don, Russia \\
\it $^2$Institute of Astronomy, University of Vienna, Austria \\
\it $^3$SHARCNET/UOIT Consortium, Oshawa, Canada \\
\it $^4$Physics Department, Southern Federal University, Rostov-on-Don, Russia \\
\it $^5$Special Astrophysical Observatory, Russian Academy of Sciences, \\
Nizhnyi Arkhyz, Karachay-Cherkessia Republic, Russia}

\date{}

\maketitle

\begin{abstract}
We consider the evolution of the ionization zone around Population III stars with $M_*\sim 25-200~M_\odot$
in protogalaxies with $M\sim 10^7~M_\odot$ at redshifts $z = 12$, assuming that the dark matter profile is
a modified isothermal sphere. We study the conditions for the growth of instabilities in the ionization zones.
The Rayleigh-Taylor and thermal instabilities develop efficiently in the ionization zones around 25-40~$M_\odot$
stars, while this efficiency is lower for stars with $\sim 120~M_\odot$. For more massive stars ($\sim 200~M_\odot$),
the flux of ionizing photons is strong enough to considerably reduce the gas density in the ionization zone, and the
typical lifetimes of stars ($\sim 2$~Myr) are insufficient for the growth of instabilities. The gas in a protogalaxy
with $M\sim 10^7~M_\odot$ with a 200~$M_\odot$ central star is completely ionized by the end of the star's lifetime; 
in the case of a 120~$M_\odot$ central star, only one-third of the total mass of gas is ionized. Thus, ionizing photons
from stars with $M_*\simlt 120~M_\odot$ cannot leave protogalaxies with $M\simgt 10^7~M_\odot$. If the masses of the
central stars are 25 and 40~$M_\odot$, the gas in protogalaxies of this mass remains essentially neutral. We discuss the
consequences of the evolution of the ionization zones for the propagation of the envelope after the supernova
explosions of the stars and the efficiency of enrichment of the intergalactic medium in heavy elements.
\end{abstract}



\section{Introduction}

\noindent

Various instabilities can arise in the ionization
zones around massive stars, related to the thermodynamics
of the gas behind the ionization front and/or
shock, as well as the break-out of the ionization
front [1-3]. These instabilities lead to substantial
fluctuations of the density and temperature and increase
the velocity dispersion of the gas in the ambient
interstellar medium (ISM). Hence, subsequent
supernovae will occur in the medium with random
flows, probably facilitating a rapid break-up of the supernovae
envelopes and increasing the efficiency of
mixing of heavy elements in the ISM [4]. These
effects are not only important for the ISM [5], but are
especially interesting for the problem of the primordial
enrichment of the first protogalaxies by Population III
stars.

The evolution of ionization zones around the first
stars has been studied in some detail [6-9]. The
main focus of these papers is the global dynamics
of the propagation of the ionization front and the
influence of the ionizing radiation of the first stars
on the efficiency of photo-evaporation of gas in the
first protogalaxies, or in other words, on their destruction
[10]. Another problem was the calculation of the
number of ionizing photons leaving the protogalaxy,
or the escape fraction [11, 12], which is important for
studies of the possibility of observing the ionization
zones [13] and the dynamics of the ionization of the
intergalactic medium [14], and, hence, reionization of
the Universe. It is evident that more accurate studies
of the evolution of the propagation of the ionization
fronts require investigation of the dynamics on small
scales and the possible growth of instabilities in ionization
zones around the first stars. The first attempt
to investigate this problem using three-dimensional
modeling was carried out in [15], which assumed
that the initial density profile in the cloud followed
a power-law, $\sim r^{-2}$ (steeper distributions were also
studied, with the aim of investigating the growth of
the Rayleigh-Taylor instability); to some extent, this
assumption corresponds to the conditions in the first
protogalaxies. However, note that the gas profile also
depends on the dark-matter distribution in the protogalaxy,
and more realistic studies of the efficiency
of the growth of instabilities in the ionization zones
taking into account the dynamics of gas cooling in the
dark-matter potential are necessary.

Here, we study the growth of instabilities in
ionization zones around the first massive stars in
spheroidal (non-rotating) dwarf protogalaxies with
the dark-matter density profile corresponding to a
modified isothermal sphere. We used a cosmological
model with a $\Lambda$-term and cold dark matter ($\Lambda$CDM
model) with $(\Omega_0,\Omega_{\Lambda},\Omega_m,\Omega_b,h ) 
= (1.0,\ 0.76,\ 0.24,\ 0.041,\ 0.73 )$, and assumed a relative abundance of
deuterium $n[{\rm D}]/n = 2.78\cdot 10^{-5}$ [16].


\section{Model of a protogalaxy and numerical methods}

\noindent

This section presents a brief description of the
main model parameters and numerical methods applied
(more detail may be found in [17, 19]), as well as
the initial conditions used.

\subsection{Main Parameters}

In the model considered, the protogalaxy consists
of gas surrounded by a spherically symmetric halo of
dark matter. The dark-matter density profile corresponds
to a modified isothermal sphere:
\be
 \rho_h(r) = {\rho_0 \over 1 + (r/r_0)^2},
 \label{halo}
\ee
where $r_0$ and $\rho_0$ are the radius of the nucleus and
the central density, respectively. We assumed a total
mass of the protogalaxy (dark halo and gas) of $M_{\rm h} = 
10^7~M_\odot$; for redshift $z = 12$, this corresponds to $3\sigma$
perturbations in a $\Lambda$CDM model with the parameters
derived from the three-year observations of the
CMB radiation by the WMAP satellite [16]. Such a
protogalaxy has a virial radius $r_{\rm v} = 520$~pc (for virial
relations, see, for instance, [20]). Simple estimates
[21, 22] indicate that $10^7~M_\odot$ protogalaxies at redshift
$z = 12$ cooled efficiently and produced the conditions
necessary for the birth of first-generation stars. Because
of the low efficiency of the fragmentation of
the primary gas, we assume that in the low-mass
protogalaxies initially appears only one star (see, e.g.,
[20]), although under certain conditions, in particular
in rotating protogalaxies, there may appear a group of
stars [19].

Since the minimum temperature of the primordial
gas at large redshifts varies from 40 to 200 K (it is
determined by the efficiency of the formation of H$_2$
and HD molecules [21]), the accretion rates onto
protostellar cores were higher than for the modern
elemental abundances ($\dot M \sim c_s^3$). Therefore, first
generation stars were much more massive than stars
of subsequent generations. Numerical models suggest
that their masses varied over a wide range of
$\sim 10-10^3~M_\odot$ (see, for instance, [20]). The massive
stars formed from the primordial gas were much
brighter than stars of subsequent generations containing
heavy elements, since the $p-p$ chain requires
higher temperatures than the CNO cycle. Of the total
mass range, we shall study the ionization zones for
25, 40, 120, and 200~$\msun$ stars. We made this choice
because these stars have appreciably different characteristics
(see the Table), and such stars ultimately
explode as supernovae, but do not collapse into black
holes, as is the case with the stars $M > 260~\msun$ [23].

\begin{table}[!ht]
\caption{Main characteristics of the first stars \cite{schaerer}.}
\center
\begin{tabular}{ccc}
\hline
\hline
   Mass, $M_\odot$  & Lifetime, Myr &    $\dot N_H$, s$^{-1}$  \\
\hline
25   &   6.459   &  $7.583\times10^{48}$    \\
40   &   3.864   &  $2.469\times10^{49}$    \\
120  &   2.521   &  $1.391\times10^{50}$    \\
200  &   2.204   &  $2.624\times10^{50}$    \\
\hline
\hline
\end{tabular}%
\label{table1}
\end{table}

\subsection{Numerical Methods}

The gas dynamics in the protogalaxy model considered
is described using the usual gas-dynamical
equations in cylindrical coordinates in the approximation
of axial symmetry ($z,r,\phi$), which were solved
numerically using a finite-difference method with an
operator splitting technique [25]. A piecewise-parabolic,
third-order interpolation scheme was used for the gas
transport [26].

To obtain the initial distribution of the gas density
in the external gravitational potential of the dark matter
$\Phi_{h}$, the equilibrium equations in cylindrical coordinates
$(z, r)$ were solved numerically. The gas was
assumed to be neutral, to have molecular mass $\mu = 1.22$, 
and to be isothermal, with the temperature equal
to the virial value $T_{\rm vir}(M)$. To solve the equilibrium
equations, it is necessary to know the initial velocity of
the gas $u_{\rm \phi0}$, which we obtain from the relation 
$u_{\rm \phi 0}=\alpha \, u_{\rm circ}$, where $u_{\rm circ}$ 
is so-called circular velocity of the
gas, which is determined by the gravitational potential
of the dark matter only (see the virial relation,
for example, in [20]), and $\alpha$ is a parameter smaller
than unity. The method used to solve the equilibrium
equations is described in [27]. The iterations were
continued until a gas-density profile corresponding to
the mass of gas inside the virial radius was obtained,
$M_{\rm g}= (\Omega_b / \Omega_m ) M_{\rm h}$. We restricted our consideration
to spheroidal (non-rotating) protogalaxies.

\subsection{Chemical Kinetics and Radiative Transfer}

It was assumed in our model that all chemical
agents (atoms, ions, molecules) are passive, i.e., they
have a velocity field similar to that of the gas. This
makes it possible to consider the gas motion only.
The chemical kinetics of the primordial gas takes
into account the following main components: 
H, H$^+$, H$^-$, He, He$^+$, He$^{++}$, H$_2$, H$_2^+$, D, D$^+$ and HD.
We computed the electron number density by assuming
charge conservation. The helium abundance (by
mass) was set to $Y_{\rm He} = 0.24$. The rates of chemical
reactions for collisional and radiative processes were
taken from [28]. The chemical-kinetics equations
were solved using the backward differencing scheme~[29].

We computed the radiative transfer of the ionizing
photons using a grid of radial rays with their origin at
the center of the protogalaxy. The number of rays was
chosen such that each cell of the two-dimensional
grid was crossed by at least 10 rays. In contrast to the
three-dimensional case, it was not necessary to split
the rays as the distance to the central source grows
for the homogeneous two-dimensional grid. The ionizing
luminosity was isotropically distributed over all
rays at the center of the protogalaxy. In each cell, we
summed the number of absorbed photons along each
radial ray that crosses the cell, then used this number
of photons to compute the rates of photochemical
reactions and the corresponding heating rate of the
gas. The radiative transfer was computed for three
spectral ranges: from the threshold for hydrogen ionization
to the threshold for single ionization of helium
(13.6-24.6~eV), from the threshold for single ionization
of helium to the threshold for double ionization
of helium (24.6-54.4~eV), and from the threshold for
double ionization of helium to infinity. The spectral
integration assumed that the spectrum of the central
star did not change during the transfer process within
each of the three spectral ranges, though the relative
weights of these ranges changes during the computations.

The method we used was tested many times, both
via comparison with the exact results of simple tests
for the velocity of propagation of the ionization front
in the spherically symmetric case, and via comparison
with other codes for cosmology applications [18].

\subsection{Initial Conditions}

To improve the resolution at the center of the
protogalaxy, we applied a nonuniform computational
grid [19]. The grid spacing can be controled using a
coefficient A: reducing A increases the resolution in
inner regions of the computational grid and reduces
the resolution in outer regions. We used $A = 1.5$ and
a 900$\times$900 grid, which provides a physical resolution
better than 0.1 pc in the inner 10 pc, which decreases
to $\sim$1 pc at $r \simeq 200$ pc. The nonuniform spacing is
similar along both $z$ and $r$.

Our computations assumed that the dark matter
in the protogalaxy at redshift $z = 12$ was already virialized,
and that its configuration can already be described
by (1). Later, the gas in the protogalaxy cools
and the concentration of gas in the central region
grows, becoming $\sim 10^8$~cm$^{-3}$ in our computations.
Since physical processes that we did not take into
account in our model become important at higher gas
concentrations, we assumed that a star is born at the
center of the protogalaxy when above number density
is attained


\section{Numerical computations}

\noindent

Let us consider the evolution of the ionization
zones around first stars with masses of 25, 40, 120 and 200~$M_\odot$ 
in spheroidal (non-rotating) protogalaxies
with total masses of $10^7M_\odot$ (including dark and
barionic matter) that are virialized by redshift $z = 12$.
Figure 1 shows the distributions of the gas density
and temperature around 25, 40, 120 and 200~$M_\odot$
stars in the central region of a protogalaxy at $t = 0.99 t_{lf}^*$, 
where $t_{lf}^*$ is the main-sequence lifetime of
the star (Table 1, [24]). It is clear that the distribution
of the gas in the vicinities of the massive stars
depends strongly on the mass of the star. A 200~$\msun$
star ionizes and heats the gas not only in the region
$r < 0.1$~kpc, but also outside the virial radius for a $10^7M_\odot$
protogalaxy, as can be seen in the upper part of Fig. 2.
Small fluctuations are visible in the distributions of
the density and temperature inside the ionization zone
of such a star, and the average temperature of the
gas is  $\sim 3\times 10^4$~K. The ionization zone for a lower
mass star (120~$\msun$) is much smaller, $r_{\rm HII}\sim 0.1$~kpc;
the fluctuations of the gas density and temperature
are somewhat higher than in the gas around a 200~$\msun$
star, but remain small. "Lobes" of hot (rarefied) and
cool (dense) gas are visible in the ionization region,
which arise due to fragmentation of the envelope
formed by the shock and subsequent compression
of these fragments by radiation. The formation of
the "lobes" is related to the shadow of the envelope
fragments [3].

\begin{figure}
\center
\includegraphics[width=15.5cm]{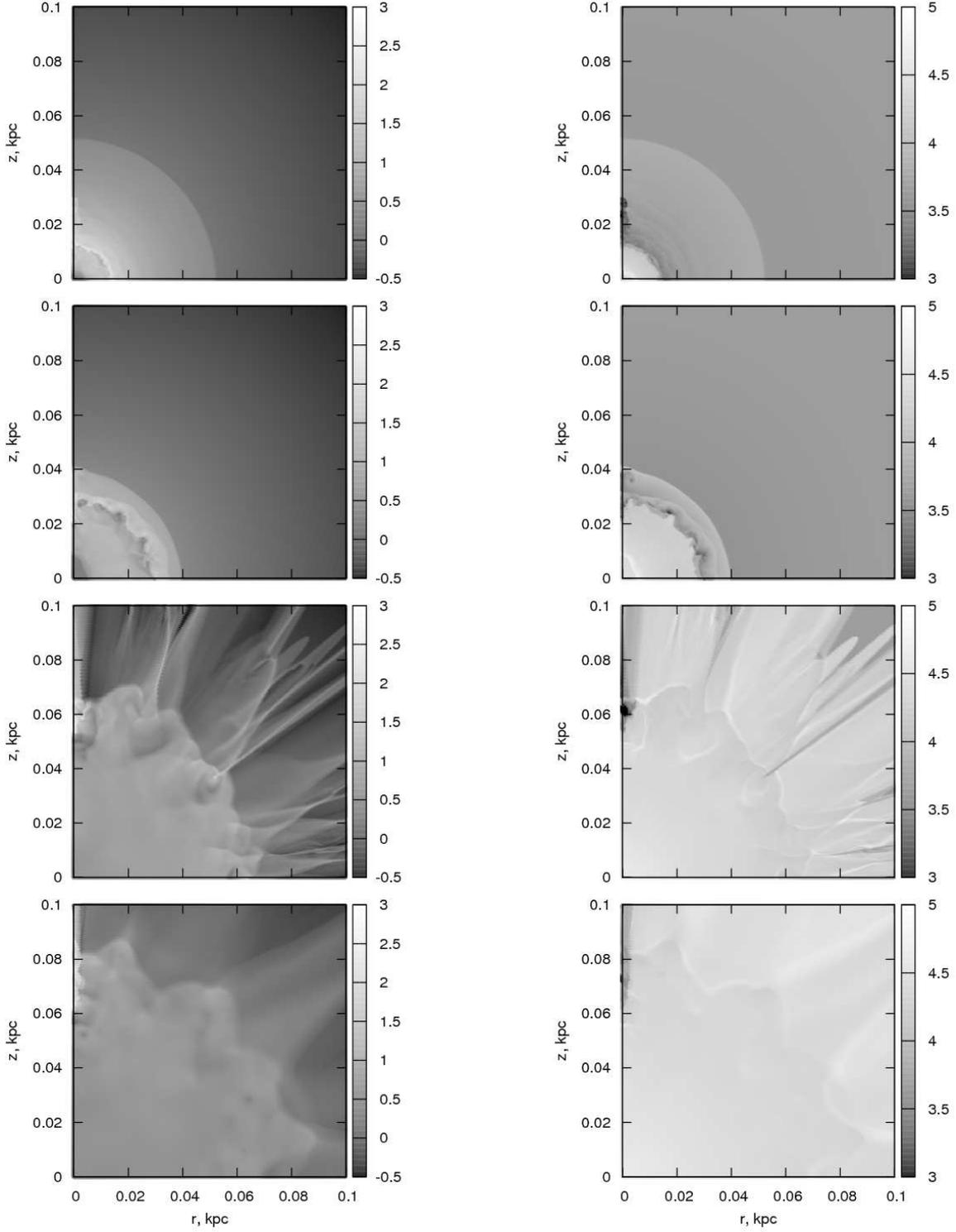}
\caption{
Maps of the distributions of the gas density (left) and temperature (right) in relative units around 25, 40, 120, and
200~$M_\odot$ stars (top to bottom) at time $t = 0.99 t_{lf}^*$, where $t_{lf}^*$ is stellar lifetime on the main
sequence (table) [24], in the central region of a spheroidal protogalaxy with total mass $M = 10^7~\msun$
(spin momentum $\lambda=0$).
}
\label{mapnt}
\end{figure}

\begin{figure}
\center
\includegraphics[width=5.5cm]{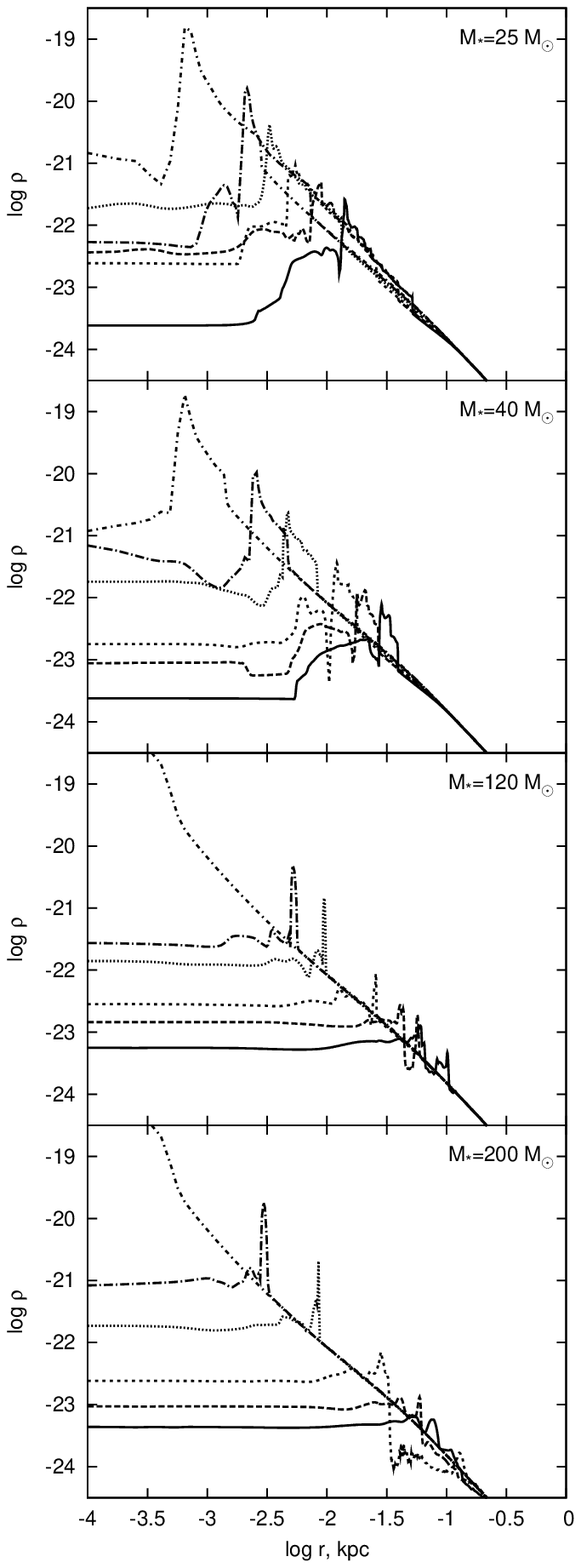}
\hspace{2cm}
\includegraphics[width=5.5cm]{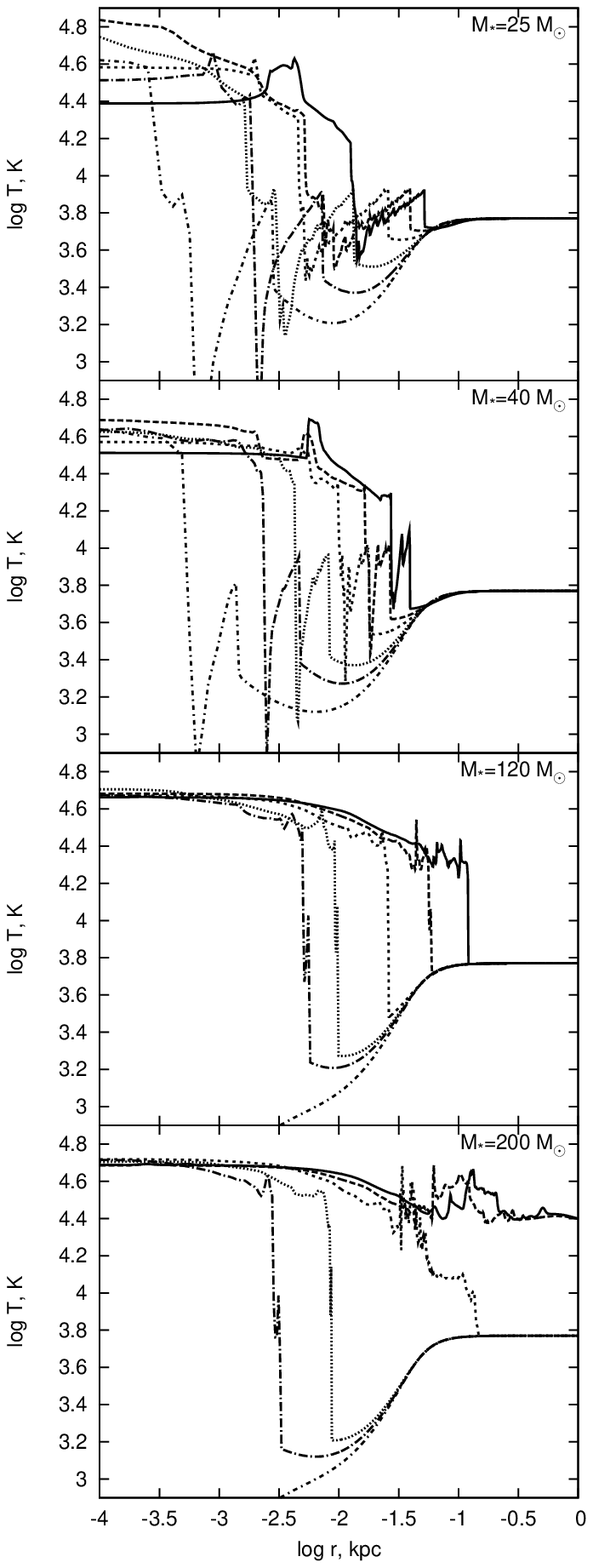}
\caption{
Radial distributions of the gas density (left) and temperature (right) around $M_*$ = 25, 40, 120, and 200~$M_\odot$ 
stars (top to bottom) along a ray from the center of a protogalaxy with total mass $M = 10^7~\msun$ at an angle 
$\pi/4$  to the vertical axis at times 0.1, 0.2, 0.3, 0.55, 0.8, 0.99~$t_{lf}^*$ (top to bottom).
}
\label{radnt}
\end{figure} 

The ionization zones for 25 and 40~$\msun$ stars are
much smaller than the virial radius of the protogalaxy:
$\sim$30 and 40 pc, respectively. At the same time, a 
D-front probably forms around such stars, in contrast
to the weak R-fronts that form around more massive
stars (120 and 200~$\msun$). Figures 1 and 2 clearly
show the ionization front itself and a distinct shock
propagating with supersonic velocity relative to the
gas around the 25 and 40~$\msun$ stars, and a propagating
ionization shock around the 120 and 200~$\msun$ stars.
The gas density behind the fronts is fairly high, and
cooling effects immediately begin to become important,
as is manifest by the enhanced density of the
gas envelope formed by the ionization front. Figure
2 shows radial profiles of the gas density and
temperature along a ray directed from the center of the
protogalaxy at an angle of $\pi/4$ relative to the vertical
axis at times $t = 0.1, 0.2, 0.3, 0.55, 0.8, 0.99~t_{lf}^*$. For
the ionization zones around the 120 and 200~$\msun$ stars,
the density jump in the envelope created by the R-front
exceeds an order of magnitude at $t\simlt 0.2~t_{lf}^*$; later, the
envelope smears out (it is disrupted). For the zones
around 25 and 40~$\msun$ stars, the profiles in Fig. 2 show
that the strong D-front formed before 0.1~$t_{lf}^*$ ($\sim$0.4 and
0.65 Myr, respectively, for 25 and 40~$\msun$ stars) and
the distribution of gas behind the front does not show
any inhomogeneities. However, at a time of  $\sim$0.3~$t_{lf}^*$,
density and temperature fluctuations arise behind the
front. These fluctuations may be related to two processes
with different natures: thermal instability and
Rayleigh-Taylor instability.

The growth of instabilities in the ionization zones
depends on the mass of the central star. This is obviously
due to the different numbers of ionizing photons
emitted by the stars. A 200~$\msun$ star emits 40 times
more photons than a 25~$\msun$ star, and the numbers
of photons emitted pairs 25 and 40~$\msun$ and 120 and
200~$\msun$ stars differ by factors of two to three (see the
table). Therefore, the nature of the instabilities in
these pairs is expected to be similar, as is seen in
Figs. 1 and 2. On one hand, low-mass stars emit
fewer ionizing photons, resulting in a lower velocity
for the ionization-front expansion than in the case of
massive stars. Thus, if the zone expands into a region
with a power-law density profile, at similar ages, the
gas density at the front of the ionization zone will be
higher for low-mass stars, making the timescales for
cooling and the growth of Rayleigh-Taylor instability
shorter. On the other hand, the lifetimes of lowmass
stars are longer, making the time available for
the growth of instabilities longer. Therefore, more
developed instability is expected in the vicinities of
lower mass stars (25-40~$\msun$), as is clearly visible in
Figs. 1 and 2.

To study variations of the physical parameters
along the ionization front, let us find the radius $r_m$
corresponding to the maximum density along a ray
directed from the center of the protogalaxy at an
angle $\pi/4$ to the vertical axis. We assume that the
circle with radius $r_m$ corresponds to the region of
the ionization front, which, in general, is close to
the real situation (Fig. 2). Next, we identify the cells
of the computational grid that are located closest to
the radius $r_m$; the polar angle $\phi$ is measured from
the vertical axis. As a result, we obtain the axial
distribution of the gas along a circle with radius $r_m$.

Figure 3 presents the distributions of the density
(lower) and temperature (upper) along circles with
radii $r_{m} = 28.6$~pc, $r_{m} + \delta r$ and $r_{m} -\delta r$, where 
$2\delta r = r_m/12$, for a 40~$\msun$ star at $t = 0.99 t_{lf}^*$.
The short-wavelength
perturbations are modulated by longer-wavelength
ones. Perturbations with different wavelengths
probably arise due to the Rayleigh-Taylor instability,
which has an increment $\gamma_{RT} \sim \sqrt{kg}$, where
$k$ is the wave vector, $g\sim \chi^{-1} v_r^2/R$ is the acceleration
of the envelope, and $\chi$ is the ratio of the densities behind
and ahead of the front. For the radius $r_{m}\simeq R = 30$~pc, 
a gas velocity $v\sim 10$~km~s$^{-1}$ and $\chi\sim 10$
the maximum wavelength for which $\gamma_{RT} t \sim 1$ at $t\simeq 4$~Myr 
is $\sim 15$~pc; this value is close to the scale of
the largest perturbations, $r_{m}\phi \sim 12-15$~pc, shown in
Fig. 2. Short-wavelength perturbations arose earlier,
when the radius of the ionized envelope was smaller.
The appearance of the smallest perturbations, with
$\lambda \simlt 1$~pc, is probably related to thermal instability,
since the ionization zone is non-stationary and radiative
heating is not completely balanced by cooling.
The cooling timescale appears to be fairly short,
$t_c \sim 10^{11}$~s, and the size of the perturbation is close
to $\lambda_t \sim c_s t_c \sim 0.03$~pc, comparable to the size of the
computational grid cells. Moreover, Rayleigh-Taylor
instability can also be significant on small scales,
since its increment grows as $\lambda^{-1/2}$.

\begin{figure}
\center
\includegraphics[width=8cm]{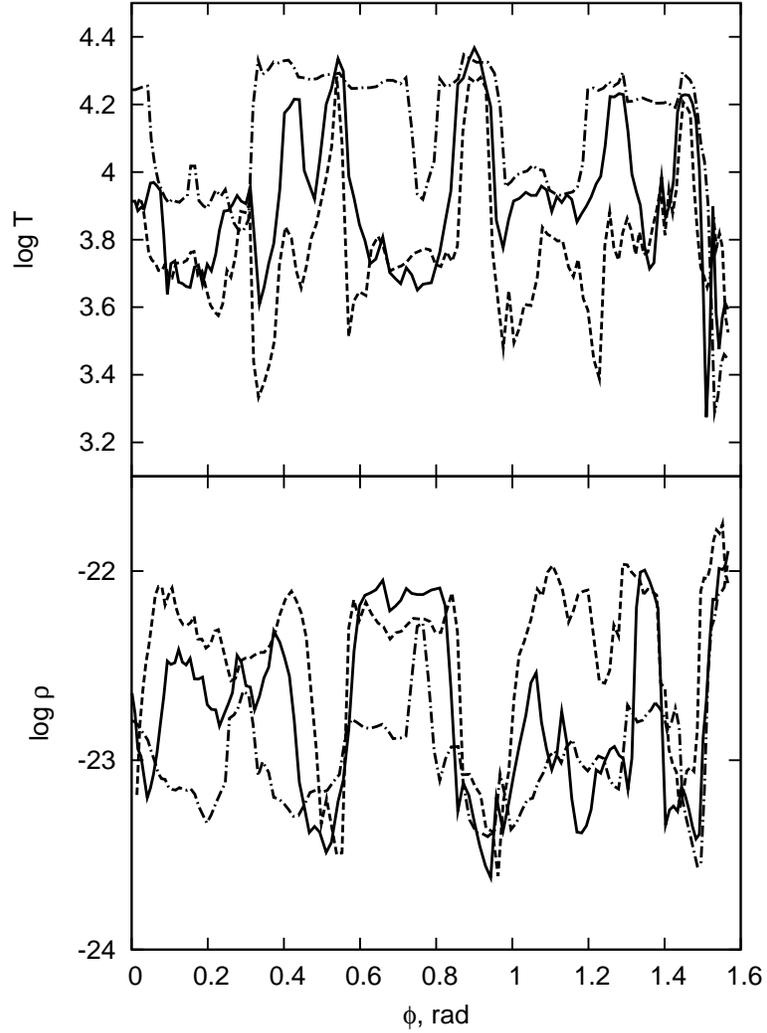}
\caption{
Distributions of the density (lower) and temperature (upper) along circles with radii of $r_{m} = 28.6$~pc
(solid), $r_{m} + \delta r$ (dashed), and $r_{m} -\delta r$ (dash.dotted), where $2\delta r = r_m/12$, for 
a 40~$M_\odot$ star at $t = 0.99 t_{lf}^*$.
}
\label{sprnt}
\end{figure} 

\begin{figure}
\center
\includegraphics[width=8cm]{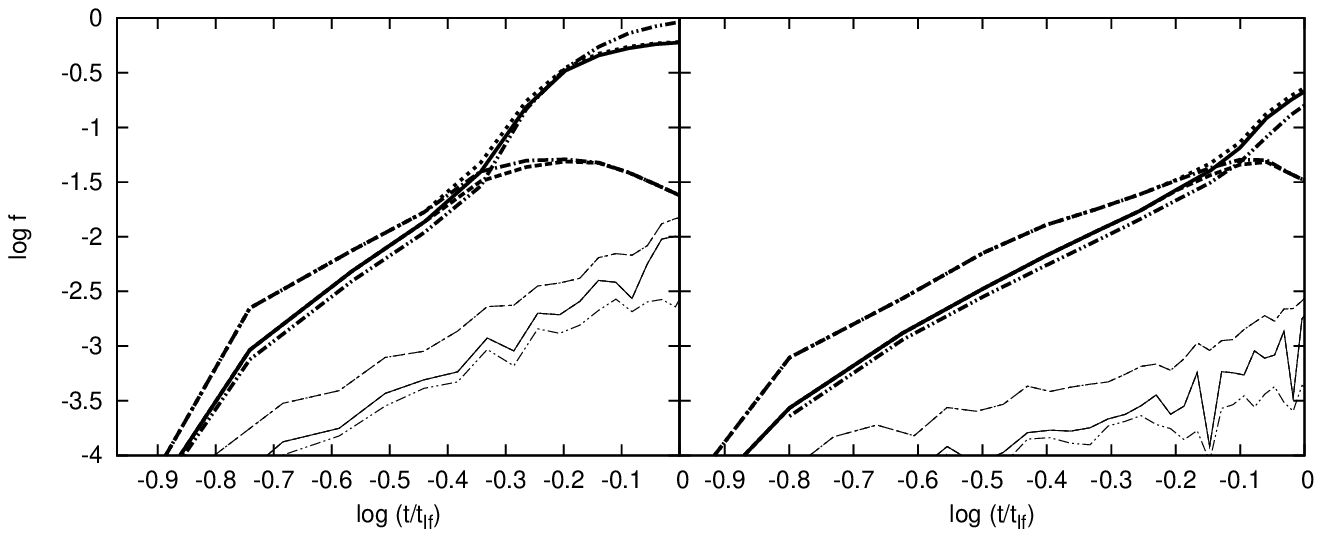}
\caption{
Evolution of the mass fraction of the gas having a relative electron number density exceeding 0.1
in the regions inside radii of $r_{vir}$ (solid) and $0.1 r_{vir}$ (dashed); mass fraction of gas 
having a relative electron number density of $10^{-3}$ in these same regions (dotted and dash-dotted, 
respectively); mass fraction of gas with temperatures higher than $2\times 10^4$~K (dash-double dotted). 
The left panel shows these dependences for 200~$M_\odot$ (thick solid) and 40~$M_\odot$ (thin) stars. The right 
panel shows the same for 120~$M_\odot$ (thick solid) and 25~$M_\odot$ (thin) stars. The curves for radii of $r_{vir}$
and $0.1~r_{vir}$ merge for the 25 and 40~$M_\odot$ stars.
}
\label{massfig}
\end{figure} 

Let now describe the global characteristics of the
ionization zones. Figure 4 shows the evolution of
the mass fraction comprised by the gas with relative
electron number density exceeding 0.1, in the region
inside radii of $r_{vir}$ (solid) and 0.1$r_{vir}$ (dashed); the
mass fraction of gas having a relative electron number
density of $10^{-3}$ in these same regions (dotted and
dash-dotted, respectively); and the mass fraction of
gas with temperatures higher than $2\times 10^4$~K (dash-
double dotted). For the 120 and 200~$\msun$ stars (thick solid),
the mass of ionized gas increases with time, and
exceeds 10\% of the mass of the protogalaxy at times
$t = 0.8~t_{lf}^*$ and 0.5~$t_{lf}^*$, respectively, while the radius of
the ionization zone exceeds 50~pc ($\sim 0.1r_{vir}$). At these
times, the degree of ionization inside the virial radius
is either less than 0.001 or greater then 0.1 (the curves
are very close), whereas the temperature of the ionized
gas exceeds $2\times 10^4$~K. The gas in a protogalaxy with
a 200~$\msun$ star is appreciably ionized at 0.8~$t_{lf}^*$, while
only one-third of the gas is ionized in a protogalaxy
with a 120~$\msun$ central star at the end of the star's
lifetime. This provides evidence that ionizing photons
from stars with $M_*\simlt 120~M_\odot$ cannot leave $M\simgt 10^7~M_\odot$
protogalaxies. If the mass of the central stars
is 25 and 40~$\msun$ (thin curves), the gas in protogalaxies
with $M\simgt 10^7~M_\odot$ remains essentially neutral, and
the ionization zone does not extend behind 0.1~$r_{vir}$.
There exists a significant fraction of the gas mass with
degrees of ionization $0.001<x_e<0.1$. It is clear that
this gas could be associated with gas that is behind
the shock, but has passed through the ionization
front. This conclusion is supported by the fact that
the mass of gas with $T> 2\times 10^4$~K is approximately
equal to the mass of gas with $x_e>0.1$ (the solid and
dash-double dot curves are very close).

To conclude, we note that the efficiency of the
ejection of heavy elements by supernovae is probably
much higher in protogalaxies with $M\simgt 10^7~M_\odot$ containing
stars with $M_*\sim 120-200~M_\odot$ than in such
protogalaxies containing stars with $M_*\simlt 40~M_\odot$,
since, first, the ionization fronts from more massive
stars substantially reduce the average density of
the gas in the protogalaxy and, second, the energy
released by the explosions of massive stars is a factor
of $\sim$3-10 higher than the corresponding energy for
less massive stars [23]. Accordingly, the radiative phase of
a supernova remnant will start earlier in protogalaxies with
lower-mass stars, and the development of Rayleigh-Taylor 
instability in the envelope is likely, so that
a significant fraction of the heavy elements will be
“locked up” in the protogalaxy as a consequence.

\section{Conclusions}

\noindent

We have considered the dynamical, thermal, and
chemical evolution of gas associated with the formation
of the ionization zones around first stars with
masses of $M_*\sim 25-200~M_\odot$ in protogalaxies with
masses of $M\sim 10^7~M_\odot$ at redshift $z = 12$, and investigated
the conditions for the development of instabilities
in the ionization zones. We have shown the
following.

\begin{enumerate}
 \item  Rayleigh-Taylor and thermal instabilities develop
in the ionization zones, which are especially
strong around 25-40~$M_\odot$ stars and less important
for stars with masses $\sim 120~M_\odot$; if the stars are more
massive ($\sim 200~M_\odot$), the flux of ionizing photons over
the entire lifetime of a star turns out to be sufficient
to create a weak ionization R-front, behind which the
instabilities do not have time to grow.
 \item The gas in a protogalaxy with $M\sim 10^7~M_\odot$ with
a 200~$M_\odot$ star is completely ionized by the end of
the star's lifetime, while only one-third of the gas is
ionized in the case of a 120~$M_\odot$ star. Thus, ionizing
photons from stars with $M_*\simlt 120~M_\odot$ are not able to
leave protogalaxies with $M\simgt 10^7~M_\odot$. If the mass of
the central star is 25 or 40~$M_\odot$, the gas in a protogalaxy
with this mass remains essentially neutral.
 \item After the supernova explosion, heavy elements
will be efficiently ejected from protogalaxies with $M\sim 10^7~M_\odot$
containing stars with masses $\sim 120-200~M_\odot$.
The ejection efficiency is negligible for stars of lower massses
(25-40~$M_\odot$); i.e., after the explosions of 25-40~$M_\odot$
stars, a significant fraction of heavy elements will be
“locked” inside the protogalaxy.
\end{enumerate}

\section{Acknowledgements}

\noindent

This work was supported by the Russian Foundation for Basic Research (project code 09-02-00933),
the Austrian Scientific Foundation (FWF, project code M 1255-N16), the Ministry of Education and
Science of the Russian Federation (the Departmental Analytical Targeted Program “The Development 
of the Potential of Higher Education” (projects RNP-2.1.1.5940, RNP 2.1.1/11879, state contract 
P-685). EOV acknowledges support from the Russian Foundation for Basic Research, project codes 
11-02-90701, 11-02-01332) and the “Dynasty” foundation. Yu.A.S. acknowledges the Russian Foundation 
for Basic Research (project code 11-02-97124). Computations were performed using computer clusters 
of the Computing Center of the Southern Federal University and the Center for Multiaccess to 
Computational Resources of the Southern Federal University. E.O.V. thanks V.N. Datsyuk for
useful discussions and advise.



\end{document}